# Determination of the hyperfine coupling constant of the cesium $7S_{1/2}$ state

Guang Yang [1, 2], Jie Wang [1, 2], Baodong Yang [1, 2], and Junmin Wang [1, 2, 3, *]

[1] Institute of Opto-Electronics, Shanxi University, Tai Yuan 030006, Shan Xi, People's Republic of China

[2] State Key Laboratory of Quantum Optics and Quantum Optics Devices, Shanxi University, Tai Yuan 030006, Shan Xi, People's Republic of China

[3] Collaborative Innovation Center of Extreme Optics, Shanxi University, Tai Yuan 030006, Shan Xi, People's Republic of China

* E-mail: wwjjmm@sxu.edu.cn

**Abstract**

We report the hyperfine splitting (HFS) measurement of the cesium (Cs) $7S_{1/2}$ state by optical-optical double-resonance spectroscopy with the Cs $6S_{1/2}$-$6P_{3/2}$-$7S_{1/2}$ (852 nm + 1470 nm) ladder-type system. The HFS frequency calibration is performed by employing a phase-type waveguide electro-optic modulator together with a stable confocal Fabry-Perot cavity. From the measured HFS between the F'' = 3 and F'' = 4 manifolds of the Cs $7S_{1/2}$ state (HFS = 2183.273 ± 0.062 MHz), we have determined the magnetic dipole hyperfine coupling constant (A = 545.818 ± 0.016 MHz), which is in good agreement with the previous work but much more precise.

## 1. Introduction

Precise measurement of the atomic hyperfine structure attracts more and more attentions for the reason that it can test the accuracy of fundamental physics. Taking the measurement of the parity non-conservation (PNC) as an example, Wood *et al* [1] measured the PNC in the $6S_{1/2}$-$7S_{1/2}$ electric dipole-forbidden transition of cesium (Cs) atoms. The PNC amplitude relies on the atomic structure calculations directly, whereas these calculations depend on the overlap between electronic and nuclear wave functions sensitively. The hyperfine structure also relies on the electron-nucleus wave functions overlap, which means that we can judge the accuracy of PNC calculations by determining the hyperfine coupling constants (HCCs) precisely [2, 3]. Moreover, precise measurement of atomic hyperfine structure can also provide more accurate benchmarks in high-precision field. Atomic transition lines which are affected by a hyperfine structure are often used as absolute frequency reference in high-resolution spectroscopy and related fundamental studies.

Hyperfine structure plays an important role in the PNC measurement, high-resolution spectroscopy, and laser cooling and trapping of atoms. However, high-precision data about the HCCs which reflect the information of hyperfine structure are still insufficient. Many groups have carried out experiments to investigate the hyperfine structure of alkali metal atoms, especially Cs and rubidium (Rb). Gupta *et al* [4] have determined HCCs about the S states of potassium (K), Rb, and Cs by cascade radio-frequency spectroscopy. Gilbert *et al* [5] have measured the hyperfine structure of the Cs $7S_{1/2}$ state by studying the directly-excited two-photon 6S$_{1/2}$-7S$_{1/2}$ transition in the presence of a strong electric field. Stalnaker *et al* [6] have used a femtosecond frequency comb to measure the absolute frequencies and the HCCs of Cs atoms. Kiran Kumar *et al* [7] have utilized the Doppler-free two-photon spectroscopy to determine the HCCs of the Cs 7D$_{3/2}$ state.

Our group has performed some measurements about the HCCs of Cs and Rb atoms. We have determined the HCCs of the Cs 8S$_{1/2}$ state [8] and the Rb 4D$_{5/2}$ state [9]. When referring to the hyperfine structure of the Cs 7S$_{1/2}$ state, it is important in testing the accuracy about the calculations of PNC. Several determinations of the Cs 7S$_{1/2}$ state have been reported over the years [4, 5], but there have been little recent research works to extend these determinations to develop a comprehensive picture of the Cs 7S$_{1/2}$ state. To reverse this situation, we have carried out the determination of the HCC of the Cs 7S$_{1/2}$ state recently. Firstly, we expect to get the high-resolution spectroscopy of the Cs 7S$_{1/2}$ state, but it is difficult to get directly through single photon electronic dipole transition which is forbidden. Two effective ways can be taken to get the spectroscopy, one is the two-photon excitation, and the other is the cascade double resonance excitation. We have chosen the latter, because the two-photon excitation is weak usually. Employing the optical-optical double-resonance (OODR) method [10, 11] via an intermediate state (the Cs 6P$_{3/2}$ state), we got the OODR spectra of the Cs 7S$_{1/2}$ state. Here, we did not use the double-resonance optical-pumping (DROP) method [12, 13], because the population of the 6S$_{1/2}$(F = 3, 4) state has not changed greatly through the 6S$_{1/2}$ (F = 3)-6P$_{3/2}$ (F' = 4)-7S$_{1/2}$ (F'' = 3, 4) and 6S$_{1/2}$ (F = 4)-6P$_{3/2}$ (F' = 3)-7S$_{1/2}$ (F'' = 3, 4) transitions. We calibrated the frequency interval by using the transmitted peaks through a confocal Fabry-Perot (CFP) cavity after the laser was phase-modulated by a fiber-pigtailed waveguide electro-optic modulator (EOM). By adjusting the length of CFP cavity and the radio frequency signal which drove the EOM, we aligned the OODR peaks with the CFP signals to reduce the nonlinearity of frequency scanning. Then we got the hyperfine splitting (HFS) of the Cs 7S$_{1/2}$ state much more precisely, and the magnetic dipole HCC was precisely determined using this method. Moreover, we also utilized the phase-modulated OODR spectra to measure the HFS of the Cs 7S$_{1/2}$ state [14].

## 2. Principles

Hyperfine structure stems from the electron-nucleus interactions. Using first-order perturbation theory, the Hamiltonian of hyperfine structure is given by [15, 16]

$$H_{hfs} = A\mathbf{I} \bullet \mathbf{J} + B\frac{3(\mathbf{I} \bullet \mathbf{J})^2 + \frac{3}{2}(\mathbf{I} \bullet \mathbf{J}) - I(I+1)J(J+1)}{2I(2I-1)J(2J-1)}, \quad (1)$$

and the eigen-energies under the hyperfine interaction could be written in terms of the hyperfine energy shift

$$\Delta E_{hfs} = \frac{1}{2}AK + B\frac{\frac{3}{2}K(K+1) - 2I(I+1)J(J+1)}{4I(2I-1)J(2J-1)}. \quad (2)$$

Here $K = F(F+1) - I(I+1) - J(J+1)$, A is the magnetic dipole HCC, B is the electric quadrupole HCC, **I** is the total nuclear angular momentum, **J** is the total electronic angular momentum, so the total atomic angular momentum **F** = **I**+**J**, and I, J, F is the quantum numbers corresponding to **I**, **J** and **F**.

For a specific state, the HFS from F to F-1 could be easily calculated as follows,

$$\Delta E_{hfs}(F \to F-1) = AF + B\frac{\frac{3}{2}F\left[F^2 - I(I+1) - J(J+1) + \frac{1}{2}\right]}{I(2I-1)J(2J-1)}. \quad (3)$$

We can infer from equation (3) that the HCCs could be determined by measuring the HFS precisely. As for the Cs 7S$_{1/2}$ state, the orbit angular momentum L = 0, which leads to the gradient of electric field outside the nucleus being zero, so there is no electric quadrupole interaction. The HFS of the Cs $7S_{1/2}$ state could be shown naturally below

$$\Delta E_{hfs}\left(7S_{1/2}, F^{"} = 4 \rightarrow F^{"} = 3\right) = A \times 4. \tag{4}$$

We can get the information of HFS through OODR spectroscopy. Usually, two beams which are corresponding to the transitions in a ladder-type atomic system are included in the OODR scheme. The OODR spectra are obtained by detecting the population difference between the intermediate state and the excited state. For the cascade Cs $6S_{1/2}$-$6P_{3/2}$-$7S_{1/2}$ transitions shown in figure 1, we can perform the OODR spectra by probing the transmission signal of the scanning probe laser L2 (1469.9 nm) when the pump laser L1 (852.3 nm) is locked. There will be five absorption peaks (corresponding to the transitions **a**, **b**, **c**, **d**, and **e** shown in figure 1 when L1 is locked to the $6S_{1/2}$ (F = 3)-$6P_{3/2}$ (F' = 4) transition) with the affection of the Doppler Effect when the HFSs of the Cs $6P_{3/2}$ state are less than the Doppler background (~1 GHz) for both counter-propagating (CTP) configuration and co-propagating (CP) configuration of two lasers L1 and L2. For the CTP configuration, the linewidth of the OODR spectra is a little bit narrow due to the atomic coherence. For the CP configuration, the frequency intervals between the nearby absorption peaks (corresponding to the transitions **a**, **b**, **c** or **d**, **e**) are wide.

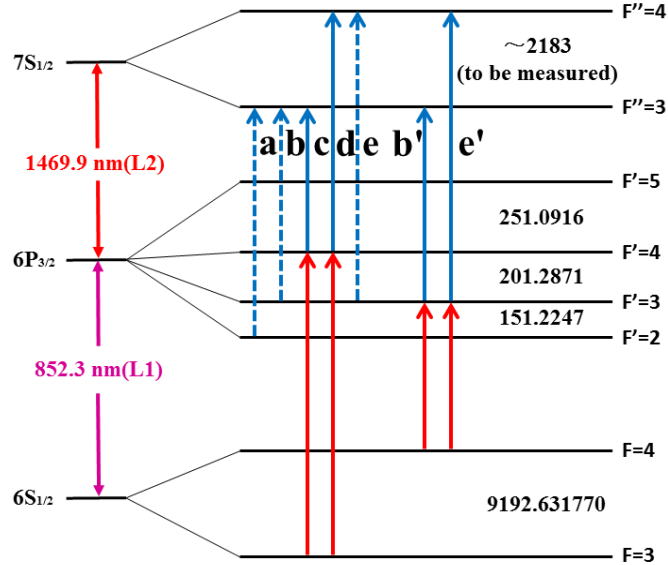

**Figure 1.** Relevant hyperfine levels of Cs atoms for the $6S_{1/2}$-$6P_{3/2}$-$7S_{1/2}$ transitions (not to scale). The numbers between the energy levels represent the numerical values of the HFS in megahertz. The $6P_{3/2}$ state values are taken from ref. [17], and the $6S_{1/2}$ state value is exact.

Taking the transitions **d** and **e** as an example, L1 is locked to the $6S_{1/2}$ (F = 3)-$6P_{3/2}$ (F' = 4) transition. But for the atoms with different velocity groups, they can be populated to not only $6P_{3/2}$ (F' = 4) manifold but also the nearby manifolds $6P_{3/2}$ (F' = 2) and $6P_{3/2}$ (F' = 3) due to the Doppler Effect. The $6P_{3/2}$ (F' = 2)-$7S_{1/2}$ (F'' = 4) transition is forbidden, so there will be five OODR spectra when we scan the L2. The $6S_{1/2}$ (F = 3)-$6P_{3/2}$ (F' = 3) transition occurs when the atoms move to L1 with the velocity v = $\lambda_1\Delta_1$, where $\Delta_1$ equals 201.2871 MHz which means the detuning of L1 relative to the $6S_{1/2}$ (F = 3)-$6P_{3/2}$ (F' = 4) transition. So the frequency intervals between spectra corresponding to the transitions **d** and **e** are $\Delta_1\lambda_1/\lambda_2$ = 116.7 MHz for CTP configuration and $\Delta_1(\lambda_1/\lambda_2 +1)$ = 318.0 MHz for CP configuration. We have chosen the CP configuration in our experiment, because it would be easy to fit the OODR spectra.

## 3. Experiment

A schematic diagram of the experimental setup is shown in figure 2. It can be divided into three sub-systems: the distributed-Bragg-reflector (DBR) type diode laser system (system Ⅰ), the external-cavity diode laser (ECDL) system (system Ⅱ), and the frequency calibration system (system Ⅲ).

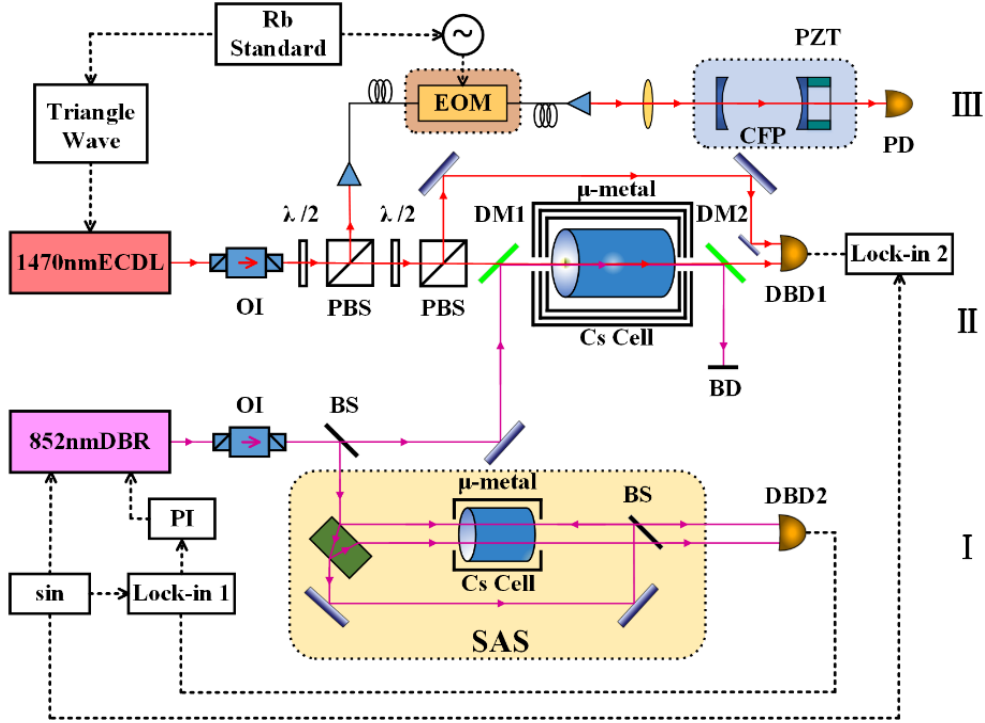

**Figure 2.** Schematic of the experimental setup. The following abbreviations are used: sin, sine-wave signal generator; PI, proportion and integration amplifier; Lock-in, lock-in amplifier; OI, optical isolator; BS, beam splitter; SAS, saturated absorption spectroscopy setup; μ-metal, high magneto-conductivity permalloy; λ/2, half-wave plate; PBS, polarization beam splitting cube; DM, 45º dichroic mirror; BD, beam dump; EOM, fiber-pigtailed waveguide-type electro-optic phase modulator; CFP, confocal Fabry-Perot cavity; Rb Standard, rubidium frequency standard; PD, photodiode; DBD, differential balanced detector; PZT, piezoelectric ceramic transducer.

The laser (L1) in system Ⅰ which acts as the coupling light is corresponding to the $6S_{1/2}$-$6P_{3/2}$ transitions. A ramp voltage provided by a function generator (Agilent 33210A) and 100 kHz sine voltage modulation signal coming from the Lock-in 1 (Stanford Research System Inc, Model SR830) were added to the current modulation input of the laser controller. Figure 3 shows the saturated absorption spectra (SAS) of the transitions. There are four channels indicated by the short dots lines shown in figure 3 we could use to lock the DBR laser for the HFS measurement. We chose the peaks $T_4$ in (a) and $T_3$ in (b) to lock the DBR laser, which was corresponding to the $6S_{1/2}$ (F = 3)-$6P_{3/2}$ (F' = 4) and $6S_{1/2}$ (F = 4)-$6P_{3/2}$ (F' = 3) hyperfine transitions. Here, we didn't chose the peaks $T_3$ in (a) and $T_4$ in (b) (they are close to the crossover line $C_{24}$ and $C_{35}$ respectively).

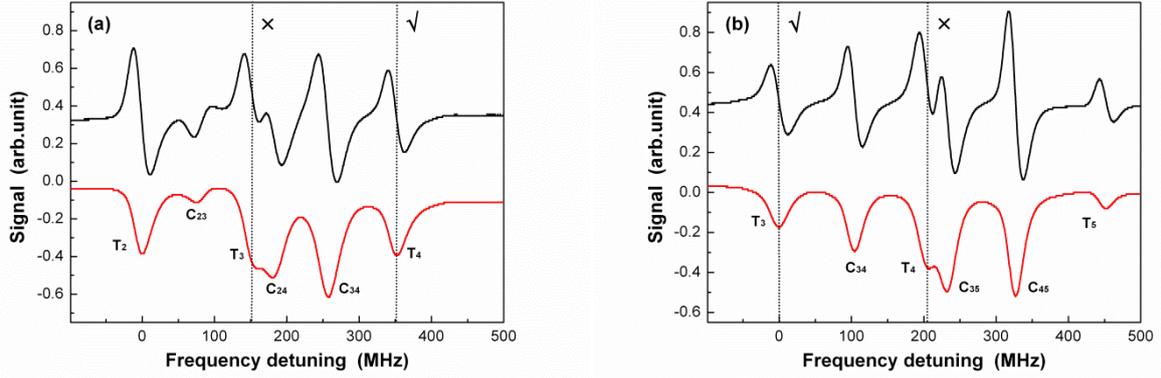

**Figure 3.** The saturated absorption spectra (the lower curve) and corresponding differential signals (the upper curve) for $6S_{1/2}$ (F = 3)-$6P_{3/2}$ (F' = 2, 3, 4) transitions (a) and $6S_{1/2}$ (F = 4)-$6P_{3/2}$ (F' = 3, 4, 5) transitions (b). The "√" alongside the short dots lines means the transition we chose for locking the laser L1. The "×" means that the transition we didn't chose.

The laser (L2) in system II acts as the probe light. The L2 overlaps with L1 in a 10-cm-long Cs vapor cell with a magnetic shielding tank (made by using of three-layer high magneto-conductivity permalloy) around. This tank reduces the magnetic field along the axis of Cs cell to less than 0.2 mG (20 nT), which is ~$10^{-3}$ less than the geomagnetic field (~500 mG). The optical power of L1 and L2 were 53 and 132 μW, the $1/e^2$ radii of the beams were 1.8 and 1.6 mm, and the polarization configuration was linear-orthogonal. Scanning the frequency of L2 while L1 was locked, we obtained the OODR spectra of the Cs $7S_{1/2}$ state. But the background of the spectra was too steep considering the intensity modulation which was led by the large frequency tuning. When we used another L2 which didn't interaction with the atoms (differential detection) as shown in figure 2 to reduce the L2's intensity modulation, we got a flat background relatively. Figure 4 shows the OODR spectra and their differential signals. The differential signals were obtained by phase sensitivity detection. The modulation frequency of L2 was 100 kHz for the reason that L2 correlated with L1 by the Cs atomic system. The frequency of reference signal produced by Lock-in 2 was also 100 kHz, because it was locked to Lock-in 1 in system I that we used for the frequency locking of L1.

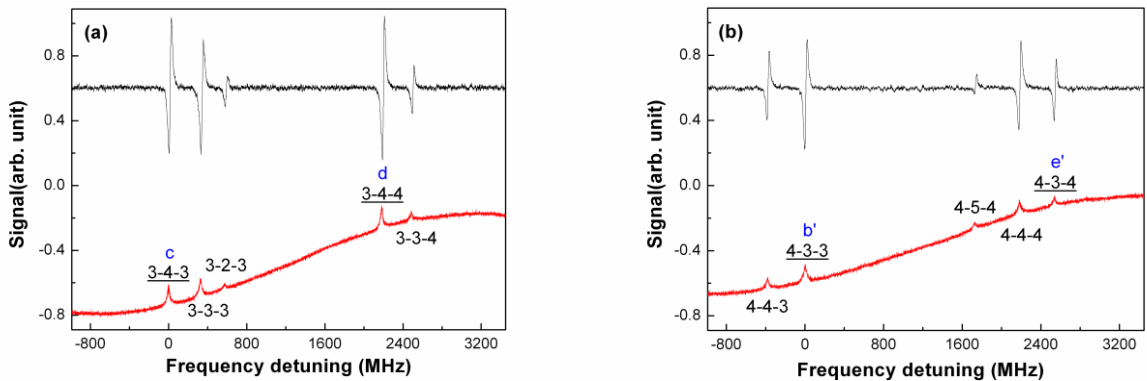

**Figure 4.** The optical-optical double-resonance (OODR) spectra (the lower curve) and their differential signals (the upper curve) for $6S_{1/2}$(F = 3)-$6P_{3/2}$(F' = 4)-$7S_{1/2}$(F'' = 3, 4) transitions (a) and $6S_{1/2}$(F = 4)-$6P_{3/2}$(F' = 3)-$7S_{1/2}$(F'' = 3, 4) transitions (b). For c, d, b' and e' transitions, please see figure 1.

System Ⅲ includes a fiber-pigtailed waveguide-type phase EOM, and a CFP cavity with a finesse of ~80 and a free spectral range of ~2.5 GHz. The EOM was driven by a frequency synthesizer (Agilent 8257D) which was locked to the rubidium frequency standard with an accuracy of ±5×10$^{-11}$ and stability <5×10$^{-12}$. The 1470 nm laser was modulated by the EOM with a radio frequency of 1090.0 MHz. By detecting the transmission of frequency-modulated laser beam, we obtained the frequency calibration signals (the CFP signals).

## 4. Results and analysis

We have got the OODR spectra and their differential signals. The differential signals and the frequency calibration signals have been chosen for the extraction of the HFS. Typical measurements are shown in figure 5, corresponding to the $6S_{1/2}$ (F = 3)-$6P_{3/2}$(F' = 4)-$7S_{1/2}$ (F'' = 3, 4) cascade transitions. The horizontal coordinate was calibrated by the 2180.0 MHz frequency interval between the 1-order sidebands in the CFP signals, which was close to the HFS of the Cs $7S_{1/2}$ state (~2183 MHz). To reduce the error brought by the frequency scanning nonlinearity of L2, we aligned the frequency calibration signals with the two-photon resonance peaks corresponding to the zero-velocity atoms among the differential signals by adjusting the CFP cavity length via the voltage driving the PZT glued on one mirror of the CFP cavity (figure 2). The frequency calibration signals and the OODR differential signals are fitted by a multipeak Voigt function and their differential form. We could see that they are an excellent fitting from the fitting residuals. After fitting the OODR differential signals and the CFP cavity signals (95% confidence level), we could get the HFS of the Cs $7S_{1/2}$ state.

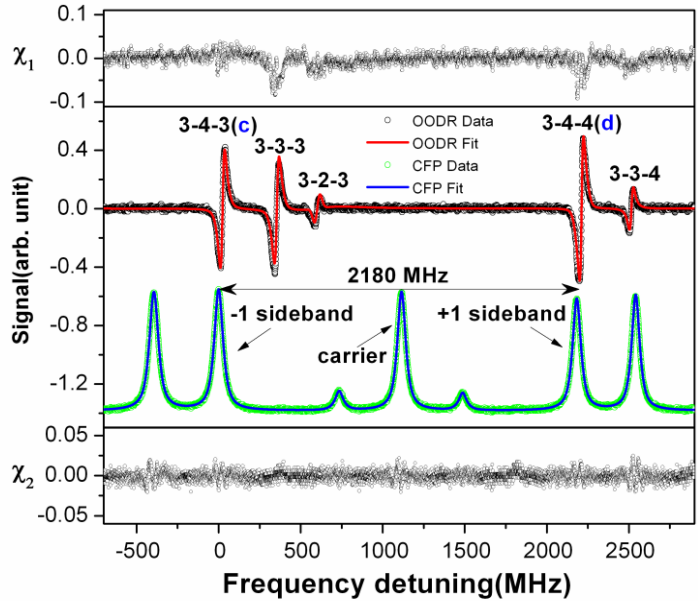

**Figure 5.** Measurement of the HFS of the Cs $7S_{1/2}$ state through the cascade $6S_{1/2}$ (F = 3)-$6P_{3/2}$ (F' = 4)-$7S_{1/2}$ (F'' = 3, 4) transitions. The middle plot: the upper curve is the differential OODR spectra, and the lower curve is the transmission signals of the CFP cavity with the scanning of L2 which is modulated by the EOM (the modulation frequency is 1090.0 MHz, therefore the frequency interval between the +1-order and -1-order sidebands is 2180.0 MHz). The small peaks between the carrier and the 1-order sidebands are the 2-order sidebands of another two cavity modes. The upper plot: residuals of the OODR fitting. The lower plot: residuals of the CFP cavity signals fitting.

For the cascade $6S_{1/2}$ (F = 3)-$6P_{3/2}$ (F' = 4)-$7S_{1/2}$ (F'' = 3, 4) and $6S_{1/2}$ (F = 4)-$6P_{3/2}$ (F' = 3)-$7S_{1/2}$ (F'' = 3, 4) transitions, the primary source of statistical error is due to the fluctuation of the laser frequency, so we recorded 40 groups of the signals. Each group included more than 100 times measurements, we fit them all (including 10360 times measurements) to acquire 40 mean HFS values and their statistical errors. And we suppose that the 40 groups have the same statistical weight. Figure 6 summarizes the experimental results of the HFS of the Cs $7S_{1/2}$ state, we obtain the mean HFS value is 2183.273 ± 0.035 MHz, where ± 0.035 MHz is the statistical error.

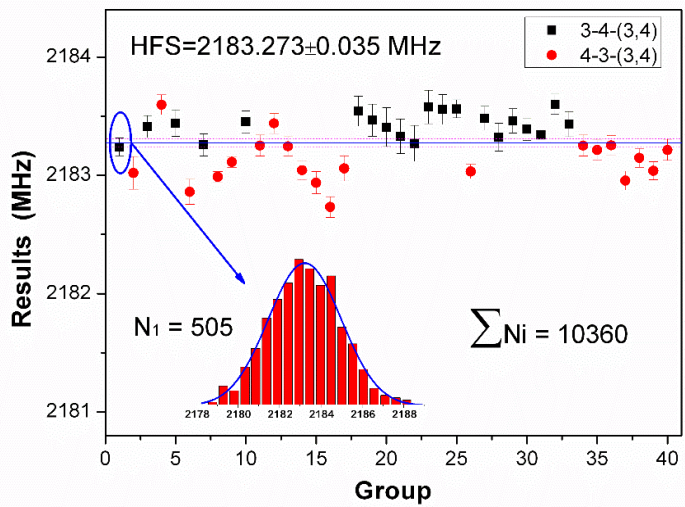

**Figure 6.** The measured HFS values of the Cs $7S_{1/2}$ state. The horizontal solid line stands for the mean value of the HFS. The range between the two horizontal dash lines stands for the statistical error. There are 505 times measurements in group 1, and the histogram of the HFS in group 1 is shown as the inset.

To precisely determine the magnetic dipole HCC, we must consider the systematic uncertainties. The uncertainty budget is summarized in table 1. The errors brought by the ac Stark shifts, the Zeeman shifts, and the pressure shifts are estimated according to our previous work [9]. Taking the ac Stark shifts as an example, we varied the power of L2 from 70 μW to 200 μW, and got the HFSs that depended on the power of L2 (group 11-17, 34-40 for the $6S_{1/2}$ (F = 4)-$6P_{3/2}$ (F' = 3)-$7S_{1/2}$ (F'' = 3, 4) transitions and group 18-24, 27-33 for the $6S_{1/2}$ (F = 3)-$6P_{3/2}$ (F' = 4)-$7S_{1/2}$ (F'' = 3, 4) transitions ). Group 11-17 and 18-24 show a systematic trend along the time line, but the ac Stark shifts are almost the same for each hyperfine manifold of the $7S_{1/2}$ state, and cause no effect on the HFS measurement because relative intervals are used. We suppose that there is another systematic error we should consider, which is the asymmetry of the differential signals.

Two reasons lead to the asymmetry of the differential signals directly in our experiment, one is that the background is not so flat although we use the differential detection, the other is that the phase of reference signal is unsuitable. The intensity modulation (the large scan range) results in the uneven background. The unsuitable phase of reference signal is probably caused by the long scan time, the weak OODR spectra, the fluctuations of lasers and many other aspects. The asymmetry of the differential signals might shift the two resonant points for the reason that we fit the signals with a symmetry form. If the relative shift of the two resonant points which is affect by the environment is positive, the result will be greater than the true value. On the contrary, the result will be smaller. In our experiment, we found that the relative shift of the two resonant points is positive for the $6S_{1/2}$ (F = 3)-$6P_{3/2}$ (F' = 4)-$7S_{1/2}$ (F'' = 3, 4) transitions as the residuals of OODR spectra shown in figure 5, and negative for the 6S$_{1/2}$ (F = 4)-6P$_{3/2}$ (F' = 3)-7S$_{1/2}$ (F'' = 3, 4) transitions. It is mainly caused by the different uneven backgrounds of these two channels as shown in figure 4. We can also see that the OODR spectra of the 6S$_{1/2}$ (F = 4)-6P$_{3/2}$ (F' = 3)-7S$_{1/2}$ (F'' = 3, 4) transitions is weaker than the 6S$_{1/2}$ (F = 3)-6P$_{3/2}$ (F' = 4)-7S$_{1/2}$ (F'' = 3, 4) transitions, so resonant points are easily shift with the fluctuations of the environment and the results of HFS measurement will also fluctuate correspondingly. We took samples from each group to judge the relative shift of the resonant points, and the error brought by the asymmetric of the differential signals is expected to be less than 50 kHz.

Misalignment of the two beams (< 2 mrad) broadens and shifts the peaks because of the first-order Doppler shift. Considering that the atomic velocity distribution is isotropic, the peaks shift in the same direction with equal distance. It can also cause a second–order Doppler shift, but the shift is so small (10$^{-1}$ kHz) [6, 7]. So we estimated the error brought by the misalignment is less than 1 kHz. Other effects, the offset of the coupling laser, the blackbody radiation, and the cell dependence are ignored because they are much smaller relatively.

**Table 1.** Uncertainty budget in measurement of the HFS of the Cs 7S$_{1/2}$ state

| Source of error | Error (kHz) |
|---|---|
| Ac Stark shifts | <5 |
| Zeeman shifts | <0.01 |
| Pressure shifts | <10 |
| Asymmetry of the differential signals | <50 |
| Misalignment of laser beams | <1 |
| Statistic error | 35 |
| Total | 62 |

So the measured HFS value of the Cs 7S$_{1/2}$ state is 2183.273 ± 0.062 MHz. Thus, we can determine the magnetic dipole HCC (A = 545.818 ± 0.016 MHz). This is in agreement with the previous values listed in table 2 but it is much more precise.

**Table 2.** HCC of the 7S$_{1/2}$ state for Cs

| Reference | A(MHz) | Method |
|---|---|---|
| Belin, 1976 [a] | 568.42 [theory] | the Fermi-Segre-Goudsmit formula |
| Dzuba, 1984 [b] | 561.51 [theory] | the RHFH method considering the correlations |
| Khetselius, 2009 [c] | 545.480 [theory] | QED perturbation theory formalism |
| Gupta, 1973 [d] | 546.3(3.0) [experiment] | Cascade radio-frequency spectroscopy |
| Gilbert, 1983 [e] | 545.90(09) [experiment] | Laser directly excited 6S$_{1/2}$→7S$_{1/2}$ transition |
| Ren, 2016 [f] | 545.93(06)[experiment] | OODR spectra with phase modulation |
| This work | 545.818(016) [experiment] | OODR spectra |

[a] The calculation mentioned in [18].

[b] The calculation mentioned in [19]. RHFH is the abbreviation of relativistic Hartree-Fock equations with the hyperfine interaction.

[c] The calculation mentioned in [20]. QED is the abbreviation of quantum electrodynamics.

[d] The measurement mentioned in [4].

[e] The measurement mentioned in [5].

[f] The measurement mentioned in [14].

## 5. Conclusion

We have determined the HCC of the Cs $7S_{1/2}$ state using the OODR spectra through the $6S_{1/2}$-$6P_{3/2}$-$7S_{1/2}$ cascade transitions with Cs vapor cell around room temperature. With the CP configuration of the coupling and probe beams, the frequency interval is larger than that of the CTP configuration, which is easy to distinguish and fit. We have calibrated the frequency axis by aligning the CFP signals with the OODR differential signals on the purpose of reducing the nonlinearity of frequency scanning. Then we got the HFS of the Cs $7S_{1/2}$ state (2183.273 ± 0.062 MHz). The final result of the magnetic dipole HCC of the Cs $7S_{1/2}$ state (A = 545.818 ± 0.016 MHz) is properly derived with considering the statistic and systematic errors. It is in agreement with the previous work [4, 5, 14], but improves the precision. It will help the theoretical study about hyperfine structure. Meanwhile, the Cs $7S_{1/2}$ state plays an important role in the PNC measurement. Dzuba et al [21] have estimated that the PNC amplitudes in the $6S_{1/2}$-$nD_{3/2}$ dipole-forbidden transitions of Cs atoms may be 4 times greater than the $6S_{1/2}$-$7S_{1/2}$ transition, but it is limited by the difficulty in handling the strong correlation effects about $n$D states. So the Cs $7S_{1/2}$ state is still significant in the PNC measurement.

**Acknowledgments**

This work is supported by the National Major Scientific Research Program of China (Grant No. 2012CB921601) and the National Natural Science Foundation of China (Grant Nos. 61475091, 11274213, and 61227902).